\documentclass[aps,prapplied,reprint,amsmath,amssymb,superscriptaddress]{revtex4-1}
\usepackage{blindtext}
\usepackage{graphicx}
\usepackage{dcolumn}
\usepackage{bm}
\usepackage{subfigure}
\usepackage{amssymb}
\usepackage{amsmath}
\usepackage{amsthm}
\usepackage{amsfonts}
\usepackage{mathtools}
\usepackage{braket}
\usepackage{soul}
\graphicspath{{./Pictures/}}
\usepackage[colorlinks,urlcolor=blue,linkcolor=blue,citecolor=blue]{hyperref}
\usepackage{xcolor}

\newtheorem*{lem*}{Lemma}
\newtheorem*{thm*}{Theorem}

\begin{document}
\title{Quantum Simulation of Polarized Light-induced Electron Transfer with A Trapped-ion Qutrit System}
\author{Ke Sun}
\email{ke.sun@duke.edu}
\affiliation{Duke Quantum Center, Duke University, Durham, NC 27701, USA}
\affiliation{Department of Physics, Duke University, Durham, North Carolina 27708, USA}
\author{Chao Fang}
\affiliation{Duke Quantum Center, Duke University, Durham, NC 27701, USA}
\affiliation{Department of Electrical and Computer Engineering, Duke University, Durham, North Carolina 27708, USA}
\author{Mingyu Kang}
\affiliation{Duke Quantum Center, Duke University, Durham, NC 27701, USA}
\affiliation{Department of Physics, Duke University, Durham, North Carolina 27708, USA}
\author{Zhendian Zhang}
\affiliation{Department of Chemistry, Duke University, Durham, North Carolina 27708, USA}
\author{Peng Zhang}
\affiliation{Department of Chemistry, Duke University, Durham, North Carolina 27708, USA}
\author{David N. Beratan}
\affiliation{Department of Physics, Duke University, Durham, North Carolina 27708, USA}
\affiliation{Department of Chemistry, Duke University, Durham, North Carolina 27708, USA}
\affiliation{Department of Biochemistry, Duke University, Durham, North Carolina 27710, USA}
\author{Kenneth R. Brown}
\affiliation{Duke Quantum Center, Duke University, Durham, NC 27701, USA}
\affiliation{Department of Physics, Duke University, Durham, North Carolina 27708, USA}
\affiliation{Department of Electrical and Computer Engineering, Duke University, Durham, North Carolina 27708, USA}
\affiliation{Department of Chemistry, Duke University, Durham, North Carolina 27708, USA}
\author{Jungsang Kim}
\email{jungsang@duke.edu}
\affiliation{Duke Quantum Center, Duke University, Durham, NC 27701, USA}
\affiliation{Department of Physics, Duke University, Durham, North Carolina 27708, USA}
\affiliation{Department of Electrical and Computer Engineering, Duke University, Durham, North Carolina 27708, USA}
\affiliation{IonQ, Inc., College Park, Maryland 20740, USA}

\begin{abstract}
Electron transfer within and between molecules is crucial in chemistry, biochemistry, and energy science. This study describes a quantum simulation method that explores the influence of light polarization on the electron transfer between two molecules. By implementing precise and coherent control among the quantum states of trapped atomic ions, we can induce quantum dynamics that mimic the electron transfer dynamics in molecules. We use $3$-level systems (qutrits), rather than traditional two-level systems (qubits) to enhance the simulation efficiency and realize high-fidelity simulations of electron transfer dynamics. We treat the quantum interference between the electron coupling pathways from a donor with two degenerate excited states to an acceptor and analyze the transfer efficiency. We also examine the potential error sources that enter the quantum simulations.  The trapped ion systems have favorable scalings with system size compared to those of classical computers, promising access to electron-transfer simulations of increasing richness.
\end{abstract}

\maketitle
Electron transfer between molecules is of central interest in energy science, signal transduction, and catalysis in both living and non-living systems~\cite{etmarcus,etdb}. Quantum effects, especially those associated with electronic coupling pathways, play a key role in the dynamics and efficiency of these reactions~\cite{prytkova2007coupling}. Light-induced electron transfer that involves many electronic and vibronic pathways can be influenced by the intensity and polarization of the excitation light~\cite{spiroscpl}. Specifically, light polarization determines the superposition of the initially prepared state. And the dynamics of electron transfer are affected by the coupling interactions mediated by interfering pathways.~\cite{ray1999asymmetric, yang2013circularly}. 

Isolating and manipulating the effects of light polarization on molecular electron transfer in experiments is challenging because of the complexity of assembling and manipulating the pathways, and also because of the dephasing interactions induced by the surroundings. Simulations are therefore widely used to study quantum dynamics. 
Trapped ion quantum simulators are proposed to offer an advantage over classical-digital simulations for issues encountered in quantum chemistry since computational resources that are intrinsically quantum mechanical in nature may be best suited for exploring quantum properties. Feynman first suggested the concept of quantum simulation in 1982, highlighting the potential for one quantum system to simulate another more efficiently than might be possible using classical computers~\cite{feynman2018simulating}. Prior works of simulating molecular quantum dynamics using trapped ions involved simulating quantum transport in a long ion chain by engineering coupling strengths based on inter-ion distances~\cite{Maier19} and simulating vibrationally-assisted energy transfer with qubits (two-level systems) and their collective motional modes~\cite{Gorman18}. 

In this Letter, we perform a quantum simulation that utilizes a fully programmable trapped-ion qutrit (three-level system) platform to simulate the light-induced electron-transfer dynamics. We employ a Trotterization method~\cite{trotter1959product, suzuki1976generalized} which enables quantum simulation of (both time-independent and dependent) Hamiltonian consisting of multiple non-commuting terms with high accuracy (see Experimental Methods for details). 

The accuracy of quantum simulations is limited, in part, by  decoherence as the system size grows~\cite{hughes1996decoherence}. A potential strategy to overcome this limitation is to use many ($d$) atomic levels per ion, or qu$d$its, when encoding the molecular Hamiltonian in the trapped-ion system~\cite{Klimov03, low2020practical, ringbauer2022universal, hrmo2022native}. Here, we use a qutrit ($d=3$) system, rather than more familiar qubit ($d=2$) structures. By using single-qutrit operations rather than two-qubit operations, we minimize the number of ions that is required for the computation, and we replace the multi-ion entangling operations with operations that manipulate the atomic levels of a single ion. This produces significantly faster operations, with a longer coherence time and higher accuracy for tracking the electron transfer dynamics being modeled (See Supporting Information for a detailed comparison between qubit and qutrit simulators). The advantages of our approach are demonstrated experimentally and analyzed using classical numerical simulations as well.\\



\noindent \textbf{Target Model.} The model for electron transfer driven by polarized light (PLET) considered here contains donor and acceptor building blocks where the electron localizes. An excitation source with adjustable polarization drives a donor-localized excitation from the ground state $\ket{G}$. Two degenerate or near-degenerate excited states ($\ket{D_1}$ and $\ket{D_2}$) are accessed. If the excited states are degenerate, the polarization of the light determines the nature of the excited-state superposition. The electronic transitions to excited states $\ket{D_1}$ and $\ket{D_2}$ are assumed to have orthogonal transition dipole moments ($\Vec{\mu}_1 \perp \Vec{\mu}_2$). Thus, the polarization of the exciting light will determine the amplitude and phase of the excited state superposition. The interaction between the degenerate donor states and the acceptor state ($\ket{A}$) is described by the couplings $V_{1}$ and $V_{2}$. 

\begin{figure}[ht]
\centering
\includegraphics[width=9cm]{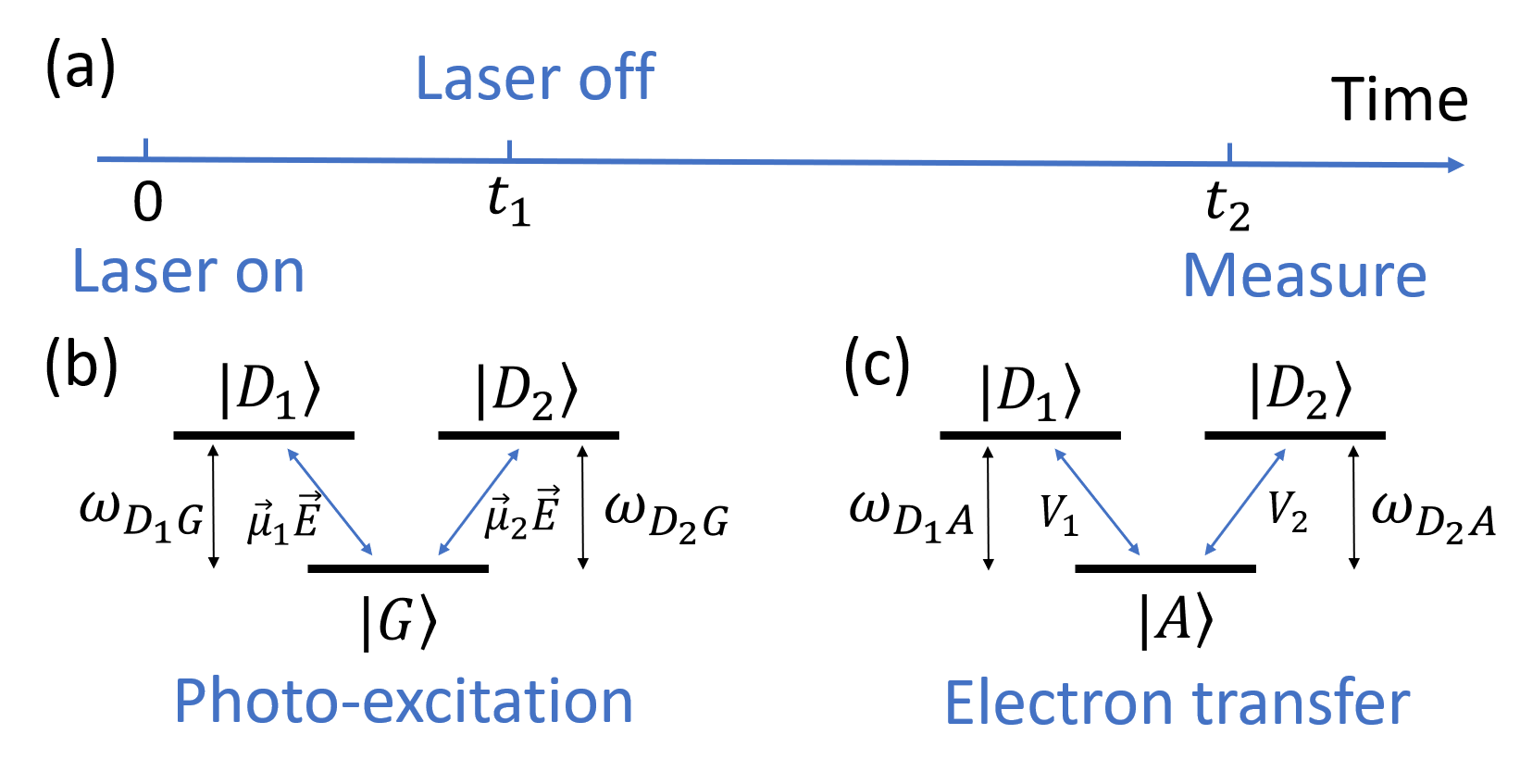}
\caption{(a) Schematic representation of the polarized light-driven electron transfer. (b) and (c) show energy level diagrams indicating the ground state, degenerate donor excited states, and the acceptor state.}
\label{fig: schematic}
\end{figure}

Electron transfer proceeds as follows (Fig.~\ref{fig: schematic}a). Exciting light impinges on the donor-acceptor (DA) system from time $t = 0$ to $t = t_1$ (Fig.~\ref{fig: schematic}b). A donor excited-state superposition, and the propagation of the electron from the donor to the acceptor, are enabled by the off-diagonal donor-acceptor couplings denoted $V_{1}$ and $V_{2}$.(Fig.~\ref{fig: schematic}c). The time-averaged acceptor population determines the electron-transfer efficiency in the time between $t_1$ and $t_2$. 

For DA systems of interest, light-driven electron transfer is much faster than radiative or non-radiative decay to the ground state. As such, the dynamics can be separated approximately into two steps. In the photo-excitation step ($t \in [0, t_1)$)  the electronic transition occurs between the ground state and degenerate excited donor states (Fig.~\ref{fig: schematic}a), since the coupling between donor and acceptor is weak. Then, in the electron transfer (ET) phase ($t \in [t_1, t_2]$), the electron migrates to the acceptor (Fig.~\ref{fig: schematic}b). Since radiative and non-radiative decay to the ground state is slower than excitation or electron transfer, only three states ($\ket{D_1}, \ket{D_2}$ and $\ket{A}$) are relevant to the dynamics.

Writing the electric field of the polarized light source (in atomic units) as $\Vec{E}(t) = [E_x(t), E_y(t), E_z(t)]$, where $x$ ($y$) is the direction of $\Vec{\mu}_1$ ($\Vec{\mu}_2$), the Hamiltonians describing the two steps are given by
\begin{align}
\hat{H}_1(t) = &\sum_j\omega_j \ket{j}\bra{j}\nonumber \\
&+ \left( \mu_1 E_x(t) \ket{G}\bra{D_1} 
+ \mu_2 E_y(t)\ket{G}\bra{D_2} + h.c.\right) \label{eqn:PLET1}\\
\hat{H}_2 = &\sum_j\omega_j \ket{j}\bra{j}\nonumber \\
&+ \left( V_{1}\ket{D_1}\bra{A} 
+ V_{2} \ket{D_2}\bra{A} + h.c. \right)  \label{eqn:PLET2}
\end{align}
where $j = \{G, D_1, D_2\}$ for the photo-excitation step and $j = \{D_1, D_2, A\}$ for the ET step, and $\omega_j$ is the energy for each state ($\hbar = 1$). \\

\noindent \textbf{Photo-excitation.} We first study the photo-excitation step, which is the first step of the PLET, described by Hamiltonian $\hat{H}_1(t)$ in Eq.~(\ref{eqn:PLET1}). Specifically, we analyze the influence of the linearly-polarized light on the electronic dynamics. 

As transition dipole moments of the two electronic transitions are orthogonal (Fig.~\ref{fig: schematic}(b)), different polarization angles will lead to different superpositions of the degenerate states in the excited-state wave function. Thus, the state of the system will have the form
\begin{equation}
    \ket{\Psi(t)} = \alpha(t) \ket{G} + \beta_1(t)\ket{D_1} + \beta_2(t)\ket{D_2}. \label{eqn: step1_state}
\end{equation}

For a linearly polarized laser, the transition dipole moment and the electric field make a constant angle $\theta$ throughout the photo-excitation process. Consequently, the ratio $r$ between the electric dipole transition strengths from the ground state to the two degenerate excited states is real and constant with time: 
\begin{equation}
    r \equiv \frac{\mu_{1}E_{x}(t)}{\mu_{2}E_{y}(t)} = \frac{\mu_1E_0\cos(\theta)\sin(\omega t)}{\mu_2E_0\sin(\theta)\sin(\omega t)} = \frac{\mu_1}{\mu_2} \cot(\theta).
\end{equation}
As a result, the ratio between the populations of the two excited states $P(\ket{D_1})$ and $P(\ket{D_2})$ is fixed over time. We quantify this ratio as the normalized population difference, defined as
\begin{equation}
    \rho\equiv\frac{P(\ket{D_1})-P(\ket{D_2})}{P(\ket{D_1})+P(\ket{D_2})}.
\end{equation} 
We also define the relative phase between the two excited states as $\phi$ where $e^{i\phi} \equiv \frac{\beta_2}{|\beta_2|}/\frac{\beta_1}{|\beta_1|}$. For linearly polarized laser, $r$ is real, so $\phi$ is either 0 or $\pi$. 

\begin{figure}[ht]
\includegraphics[width=8.5cm]{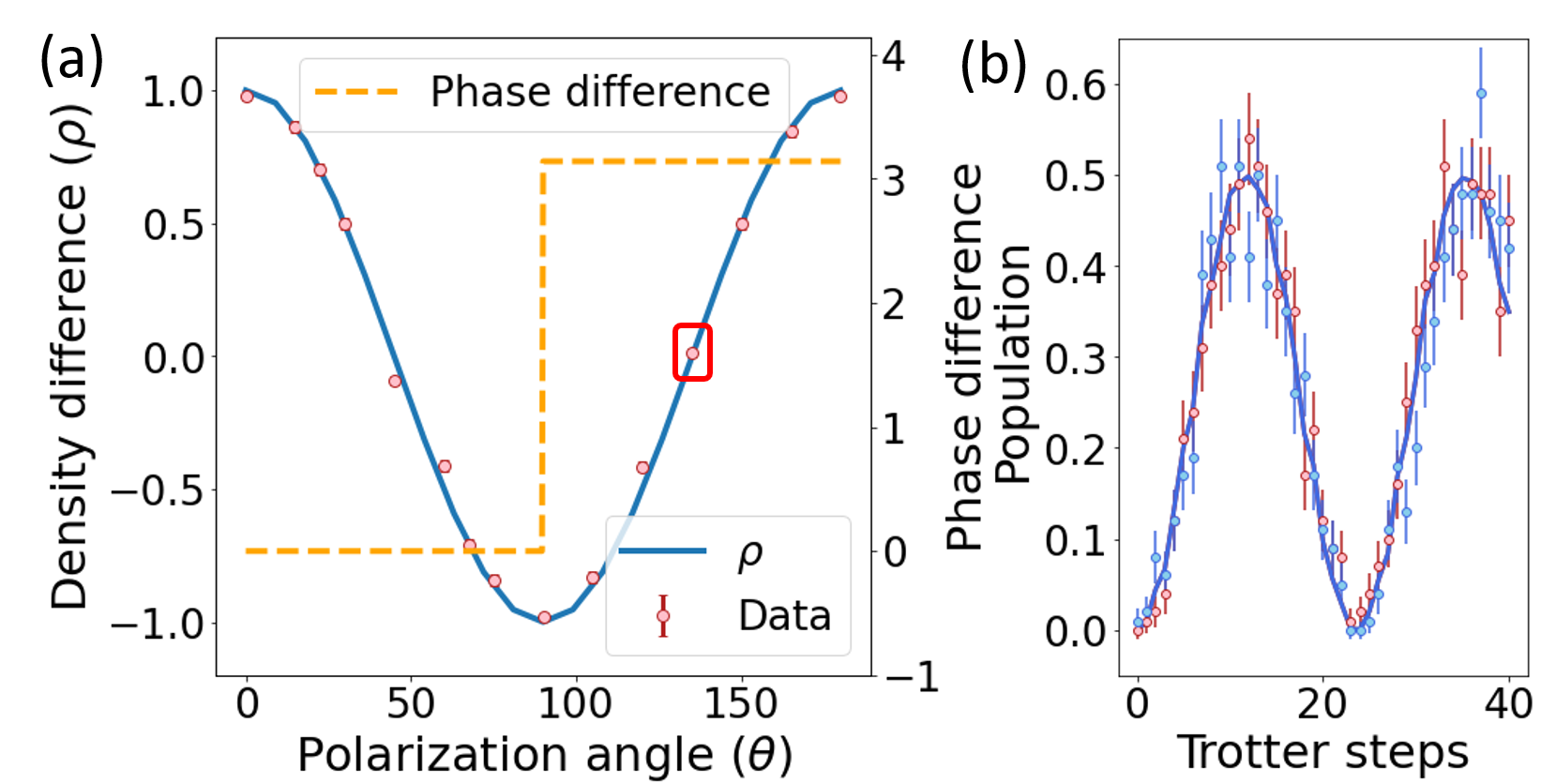}
\caption{(a) The normalized population difference and relative phase between $\ket{D_1}$ and $\ket{D_2}$, as functions of the angle of the laser light's linear polarization. (b) An example of the raw experimental data (denoted with a red rectangular box in a).  The red (blue) solid lines show the theoretical prediction of the population of $\ket{D_1} (\ket{D_2})$, while the red (blue) points represent the experimental data. This plot shows the time evolution with $E_0 = 2.2\times 10^9$ V/m, $\vec{\mu}_1 = e\cdot\{4.58,0,0\} a.u.$, $\vec{\mu}_1 = e\cdot\{0,4.58,0\} a.u.$ The donor ground and excited state energies are set to be $\omega_G = 0$, and $\omega_{D_1} = \omega_{D_2} = 3.89$ eV, respectively. The simulation is divided into 40 Trotter steps. Each step corresponds to an elapsed time of $\tau = 0.198$ fs. Scanning the Trotter steps is the same as observing the time evolution of the electron transfer. In (b), the polarization angle $\theta$ is $135^{\circ}$. The populations of $\ket{D_1}$ and $\ket{D_2}$ are the same since the projections of the electric field of the laser on both electric dipole orientations of the two excited states are the same. Thus, the solid blue and red lines overlap.}
\label{fig:step1 data}
\end{figure}

Figure.~\ref{fig:step1 data}(a) shows the normalized population difference $\rho$ and phase $\phi$ as a function of the incident polarization angle $\theta$. The population difference is obtained from both numerical calculations and quantum-simulation experiments. For the quantum simulation, we map $\ket{G}$, $\ket{D_1}$, and $\ket{D_2}$ to $\ket{0}$, $\ket{1}$, and $\ket{2}$ of the trapped-ion qutrit, respectively (see Experimental Methods). \\

\noindent \textbf{Electron transfer.} We now study the electron-transfer process, which is the second step of the PLET, described by Hamiltonian $\hat{H}_2$ in Eq.~(\ref{eqn:PLET2}). Specifically, we study the influence of (i) the relative phase of the initial states and (ii) the energy-level difference of the two excited states on the electron-transfer efficiency. 

The phase difference between the two degenerate excited states of the donor $\ket{D_{1,2}}$ determines whether the interference is constructive or destructive. The initial state is selected such that the populations of the two degenerate states are equal and the population of the ground state is zero:
\begin{equation}
    \ket{\Psi_0} = \frac{1}{\sqrt{2}}(\ket{D_1}+e^{i\phi}\ket{D_2})\label{eqn: init_state}
\end{equation}
This corresponds to $\alpha = 0$, $|\beta_1| = |\beta_2|$, and $\beta_2/\beta_1 = e^{i\phi}$ [see Eq.~(\ref{eqn: step1_state})]. 

\begin{figure*}[ht]
\centering
\includegraphics[width=16cm]{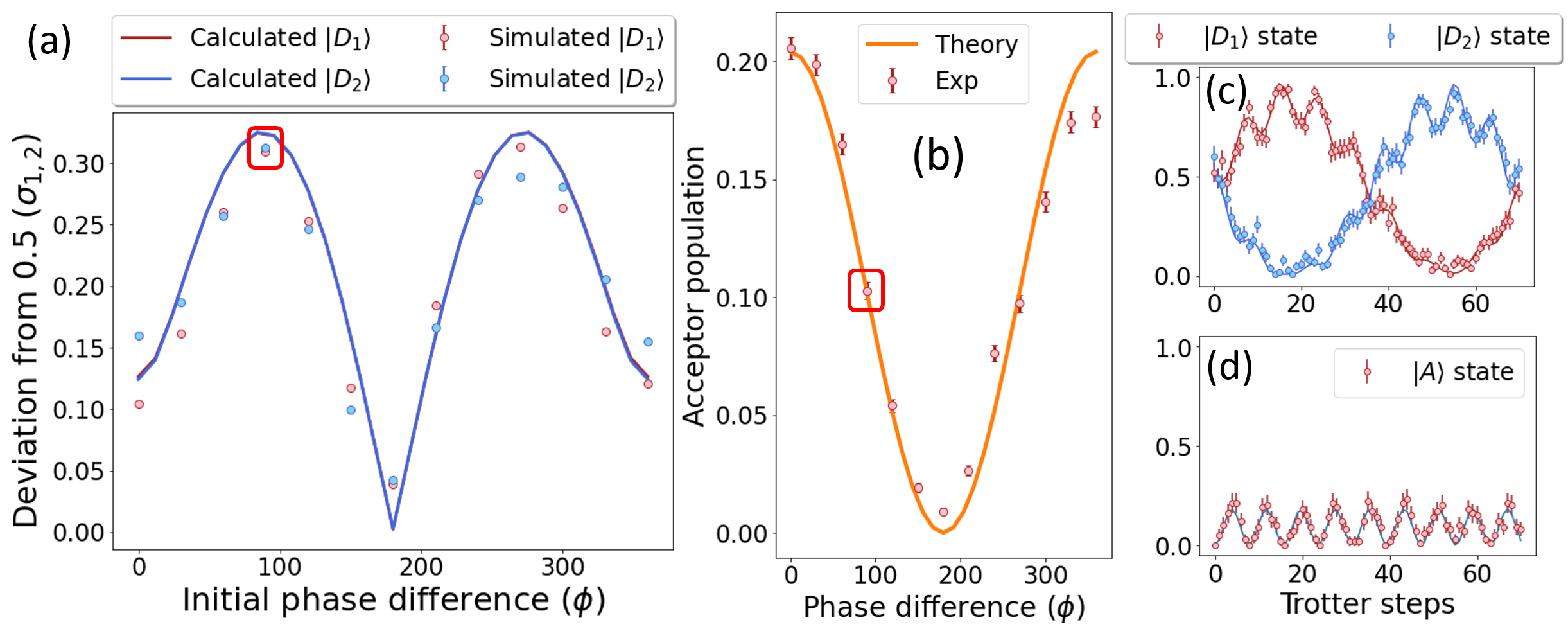}
\caption{Simulation of the electron transfer with quantum interference. The initial state is prepared as described by Eq.~(\ref{eqn: init_state}). (a) Simulated (dots) and calculated (lines) values of the time-averaged population of the two excited states as a function of the phase difference $\phi$ of the initial state, plotted as the deviation from 0.5 ($\sigma_{1,2}$). (b) The time-averaged population of the acceptor state as a function of the phase difference $\phi$ in the initial donor state. Panels (c) and (d) show the experimental (dots) and calculated (lines) time-evolution of the two donor and the acceptor states when $\phi = 90^{\circ}$ (Marked with red rectangular boxes in a and b). The horizontal axis corresponds to the number of Trotter steps used in the simulation (up to 70), each corresponding to $0.471$ fs of time evolution. Here, $\omega_{D_1} = \omega_{D_2} = 3.89$ eV, $\omega_{A} = 3.01$ eV, $V_1 = V_2 = 0.25$ eV.}
\label{fig:step2 phase scan data}
\end{figure*}

Figure~\ref{fig:step2 phase scan data} shows the results of both numerical calculations and quantum simulations describing the electron transfer to the acceptor state from this initial state. For the quantum simulation, we map $\ket{A}$, $\ket{D_1}$, and $\ket{D_2}$ to $\ket{0}$, $\ket{1}$, and $\ket{2}$ of the trapped-ion qutrit, respectively. Figure~\ref{fig:step2 phase scan data}(a) and (b) show the time-averaged values of the resulting population of the two donor states (measured as a deviation from the initial values of $|\beta_1|^2 = |\beta_2|^2 = 0.5$) and the acceptor state, respectively, as a function of the initial phase difference $\phi$. The time-averaged deviation of the donor population is defined as 
\begin{equation}
    \sigma_{1,2} \equiv \sqrt{\frac{\sum_i^N \left[P_i(\ket{D_{1,2}})-0.5\right]^2}{N}},
\end{equation}
where $i$ represents the $i^{th}$ Trotterization step. Note that when $\phi = 180^{\circ}$, the initial state does not transfer any population to the acceptor state due to destructive interference between the two coupling pathways from $\ket{D_1}$ and $\ket{D_2}$, and therefore the population of the two donor states does not change. This is also reflected by the fact that the acceptor population is zero at this value. For the parameters of the Hamiltonian summarized in the figure caption, we simulated the time evolution of the two donor state populations (Fig.~\ref{fig:step2 phase scan data}(c)) and the acceptor state population (Fig.~\ref{fig:step2 phase scan data}(d)) for an initial phase difference of $\phi = 90^{\circ}$. The horizontal axis indicates the number of Trotter steps (up to 70), representing the time evolution (each step corresponding to $\tau = 0.471$ fs of time evolution).

In the case of non-degenerate donor states $\ket{D_1}$ and $\ket{D_2}$, quantum interference is expected to be suppressed compared to the degenerate case where interference is significant. To investigate this, we begin with an initial $180^{\circ}$ phase difference that results in destructive interference and track the changes in interference as energy degeneracy is lifted.

\begin{figure*}[ht]
\centering
\includegraphics[width=16cm]{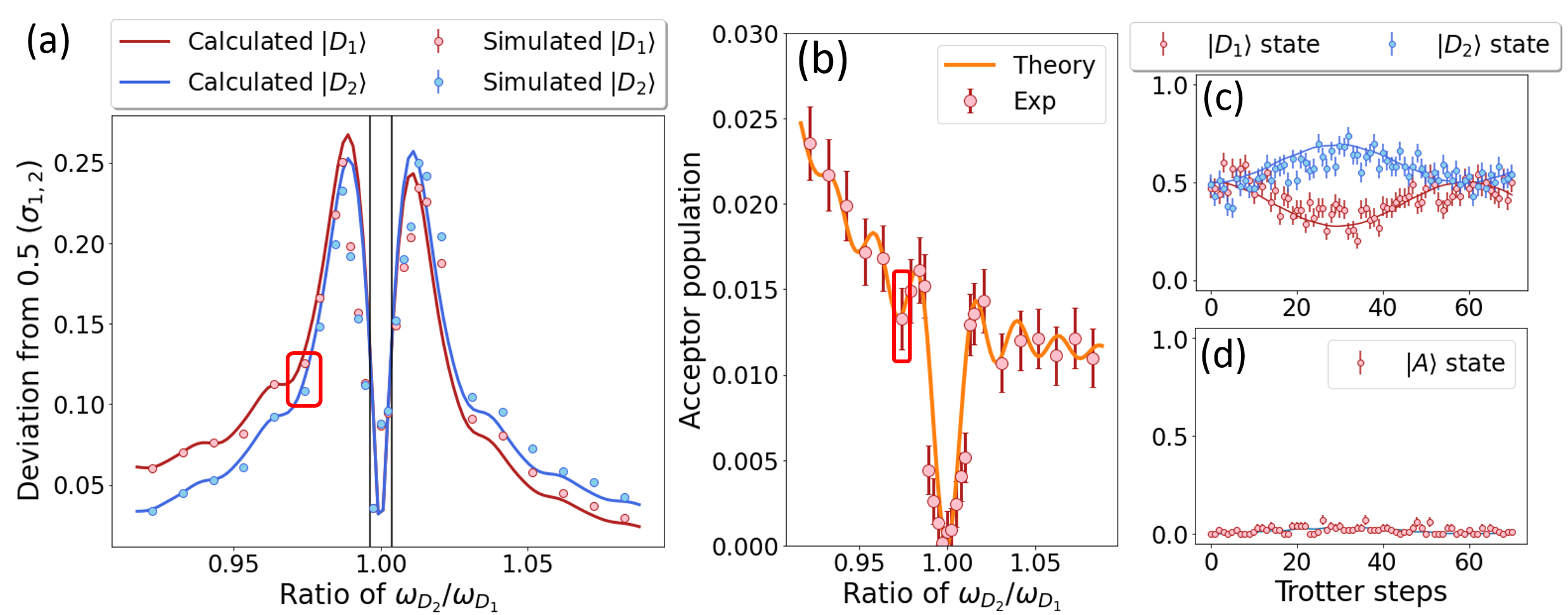}
\caption{Simulation of destructive quantum interference ($\phi = 180^{\circ}$) as donor state degeneracy is lifted. (a) Simulated (dots) and calculated (lines) values of the time-averaged population of the donor states as a function of the energy difference between the two donor states, plotted as a deviation from 0.5. (b) The time-averaged population of the corresponding acceptor state. (c) and (d) show the simulated (dots) and calculated (lines) time-evolution of the population of the two donor states and the acceptor state, respectively when $\omega_{D_2} = 3.76, \: \omega_{D_2}/\omega_{D_1} = 0.974$ (Marked with red rectangular boxes in a and b) The horizontal axis corresponds to up to 70 Trotter steps used in the simulation where each step represents $0.659$ fs of time evolution. Here, $\omega_{D_1} = 3.86$ eV, $\omega_{A} = 3.01$ eV, $V_1 = V_2 = 0.25$ eV, and $\phi = 180^{\circ}$.}
\label{fig:step2 D2 scan data}
\end{figure*}

Figure~\ref{fig:step2 D2 scan data}(a) and (b) show the time-averaged populations of the two donor states (measured as a deviation from the initial value of 0.5) and the acceptor state, respectively, as the energy $\omega_{D_2}$ of the second donor state $\ket{D_2}$ is varied from the energy $\omega_{D_1} = 3.86$ eV of the first donor state $\ket{D_1}$. For degenerate states ($\omega_{D_2}/\omega_{D_1} = 1$), the destructive interference keeps the donor state populations at 0.5 each, and the acceptor state population at 0. As the energy degeneracy is lifted, destructive interference is suppressed, with a very narrow full-width at half maximum (FWHM) linewidth of about $0.73\%$. Figures~\ref{fig:step2 D2 scan data}(c) and (d) show the quantum simulation from the ion trap system (dots) and calculated (lines) values of the donor and acceptor populations as a function of time, plotted as a function of the Trotterized steps (up to 70 steps), each step corresponding to the time evolution of $0.659$ fs, when $\omega_{D_2}/\omega_{D_1} = 0.974$. Although the acceptor state population is very small ($<0.01$) comparable to the measurement limit of our quantum simulator (determined by state preparation and measurement error), we can see clear evidence of the degradation of the destructive interference from the change in donor-state populations.\\

\noindent \textbf{Trotterization and experimental errors.} We analyze the contributions of errors to the simulations in order to determine the accuracy of our trapped-ion quantum simulator. First, the Trotterization method used in the quantum simulation inevitably introduces errors, as the time evolution of the Hamiltonian with non-commuting terms is discretized into a finite number of steps (see Eq.~(\ref{eqn:Trotterization}) of the Experimental Methods). Any implementation error arising from the experimental setup adds to this theoretical Trotterization error. Therefore, we compare the theoretically predicted state populations $P_{\rm th}$, derived from direct time evolution of the Hamiltonians in Eqs. (\ref{eqn:PLET1}) and (\ref{eqn:PLET2}), with the experimentally measured populations $P_{\rm exp}$ as well as the theoretical Trotterized predictions for populations $P_{\rm Tro}$. 

\begin{figure}[ht]
\centering
\includegraphics[width=8.5cm]{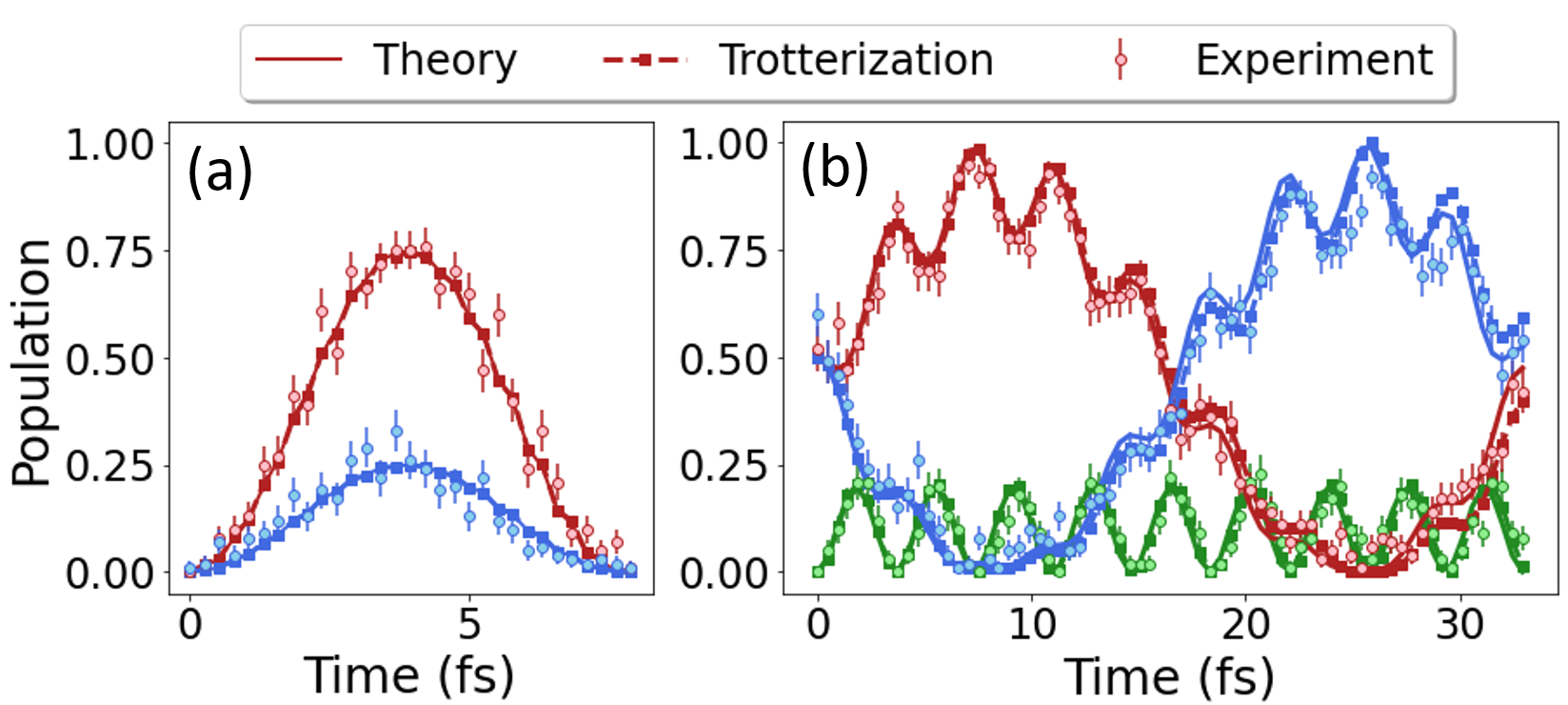}
\caption{Comparison between the theoretically calculated populations ($P_{\rm th}$, solid lines), Trotterization-method predicted populations ($P_{\rm Tro}$, dashed lines with square dots), and experimentally measured populations ($P_{\rm exp}$, dots with error bars) in (a) the photo-excitation process and (b) the electron transfer process as a function of time. In (a), populations of both excited states $\ket{D_1}$ (red) and $\ket{D_2}$ (blue) are plotted. In (b), populations of both excited states $\ket{D_1}$ (red) and $\ket{D_2}$ (blue) and the acceptor state $\ket{A}$ (green) are plotted.}
\label{fig: Data accuracy}
\end{figure}

Figure~\ref{fig: Data accuracy} shows the comparison of these three quantities. Figure~\ref{fig: Data accuracy}(a) plots the theoretical calculation of the temporal dynamics for the population of the two excited donor states $\ket{D_1}$ and $\ket{D_2}$ during the photo-excitation step as a solid line. The solid square with dashed lines shows the calculated population using Trotterization (see Eqs.~(\ref{eqn:Trotterization}) and (\ref{eqn:Trot}) of the Experimental Methods). We observe small deviations from the ideal time evolution. The circular points with error bars in Fig.~\ref{fig: Data accuracy}(a) indicate the measured values from the ion trap quantum simulator, which closely follow the theoretical values within experimental error. Figure~\ref{fig: Data accuracy}(b) shows similar comparisons for the electron transfer step for all three states $\ket{D_1}, \ket{D_2}$ and $\ket{A}$.

To analyze the deviation between the population values $P_{\rm th}, P_{\rm exp}$ and $P_{\rm Tro}$ quantitatively, we first denote the population of the state $\ket{i} (i = D_1, D_2, A)$ at the $j^{th}$ Trotterization step as $P_{x,i,j}$, where $x$ represents the ideal theoretically predicted population ($x = {\rm th}$), theoretically predicted population using Trotterization ($x = {\rm Tro}$), or experimentally measured population ($x = {\rm exp}$). We define two parameters of interest for each state $\ket{i}$: (i) the time-averaged mean distance ($\sigma_{{\rm Tro},i}$) between the population predicted by the theoretical time evolution ($P_{{\rm th},i,j}$) and the predicted population by Trotterization analysis ($P_{{\rm Tro},i,j}$) and (ii) the average mean distance ($\sigma_{{\rm exp},i}$) between $P_{{\rm th},i,j}$ and the experimentally measured population ($P_{{\rm exp},i,j}$). These two mean distances are defined as
\begin{align}
    &\sigma_{{\rm Tro},i} \equiv \sqrt{\sum_{j = 1}^{N}\frac{(P_{{\rm Tro},i,j}-P_{{\rm th}, i,j})^2}{N}}, \\
    &\sigma_{{\rm exp},i} \equiv \sqrt{\sum_{j = 1}^{N}\frac{(P_{{\rm exp},i,j}-P_{{\rm th}, i,j})^2}{N}},
\end{align}
where $N$ is the number of Trotterization steps.

In both photo-excitation and electron transfer processes, each data point in Fig.~\ref{fig:step1 data}(a), \ref{fig:step2 phase scan data}(a,b) and \ref{fig:step2 D2 scan data}(a,b) is obtained by a time evolution trial. We calculate the average distance between $P_{\rm th}$ and $P_{\rm Tro}$ for each time evolution trial and obtain $\sigma_{{\rm Tro}, i,k}$, where $i = \ket{D_1},\ket{D_2},\ket{A}$ and $k$ denotes the index of the time evolution trial. We then calculate the mean distance of each state $\Bar{\sigma}_{{\rm Tro}, i}$ vs. the theoretical value by averaging over $k$. Similarly, we compute the mean distance between the $P_{\rm th}$ and $P_{\rm exp}$ values for each state $\Bar{\sigma}_{{\rm exp}, j}$. The data are summarized in Table~\ref{tab: data accuracy}.

\begin{table}[h!]
    \begin{center}
        \begin{tabular}{| c| c | c | c | c |} 
        \hline
         {}& \multicolumn{2}{c|}{Photo-excitation} & \multicolumn{2}{c|}{Electron transfer} \\ 
        \hline
         Simulated time& \multicolumn{2}{c|}{$7.91$ fs} & \multicolumn{2}{c|}{$32.91$ fs} \\ 
        \hline
        States & $\Bar{\sigma}_{\rm Tro}$ & $\Bar{\sigma}_{\rm exp}$ & $\Bar{\sigma}_{\rm Tro}$ & $\Bar{\sigma}_{\rm exp}$ \\ 
        \hline
        $\ket{D_1}$ &  0.0067(18) & 0.043(9) & 0.024(3) & 0.071(13) \\
        \hline
        $\ket{D_2}$& 0.0051(15) & 0.044(9) & 0.024(3) & 0.078(13)\\
        \hline
        $\ket{A}$& - & - & 0.018(4) & 0.043(9) \\
        \hline 
        \end{tabular}
        \caption{Mean distances of the state populations compared to the ideal theoretical values for both photo-excitation and electron transfer processes.}\label{tab: data accuracy}
    \end{center}
\end{table}
The electron transfer process exhibits a larger $\Bar{\sigma}_{\rm Tro}$ than the photo-excitation process due to the accumulation of error caused by the Trotter approximation over increasing simulation time. To mitigate this, increasing the number of Trotter steps to decrease the value of $T/N$ in Eq.~(\ref{eqn:Trotterization}) is a straightforward solution. However, due to experimental noise, the measured $\Bar{\sigma}_{\rm exp}$ is not dominated by Trotter error. Improving qutrit coherence time and accuracy of state preparation and measurements can reduce the experimental error. If experimental noise is no longer a dominant source of error, increasing $N$ needs to be considered.\\

\noindent \textbf{Conclusion and outlook.} In this study, we utilized a trapped-ion qutrit to simulate electron-transfer dynamics in a model molecular system as a function of driving light polarization. We find that the photo-excitation process and the population difference for the two degenerate states are influenced by the exciting light polarization. We also studied the influence of phase differences between the two electron transfer coupling pathways on the electron transfer efficiency, finding that destructive interference only occurs when the energy level difference is very small (less than or equal to $0.73\%$). Quantum simulation of the few-state orbital model system can be performed well on classical-digital computers. However, the mapping of this model onto a single qutrit with low gate errors offers an encouraging approach to simulate more complex quantum systems using trapped-ion qudits, potentially exceeding the current capabilities of classical-digital computers.

Future research directions include simulating electron and excitation transfer in molecular systems with more complex connectivity, such as chemically linked donor-acceptor assemblies, which have a larger number of states. Such studies will likely require entangling operations on more than one ion, even with the use of qudits instead of qubits. The coherent control of trapped-ion qudits is a promising avenue for achieving more efficient and accurate simulations, potentially requiring fewer ions and entangling operations than qubits. However, two-qudit gates with higher fidelity are needed to conduct more complex simulations, as highlighted in recent studies~\cite{Klimov03, low2020practical, ringbauer2022universal, hrmo2022native, chi2022programmable, cervera2022experimental}.

Another interesting future direction is to add interactions between the electronic states, and between these states and the surrounding environment~\cite{Potocnik18, Wang18, Maier19}. For the PLET model, it is expected that the effects of light polarization will be diminished when interactions with the environment reduce electronic coherence. To study this behavior, we can simulate PLET structures that have various interactions with the surrounding environment. The environment is often modeled as a bath of harmonic oscillators, which can be mapped to the normal modes of the trapped ions' motion~\cite{MacDonell21}. Indeed, a donor-acceptor system coupled to one or two harmonic oscillators has been simulated in recent experiments with trapped ions~\cite{Gorman18, whitlow2022simulating, valahu2022direct} and superconducting qubits coupled to electromagnetic cavities~\cite{wang2023observation}.

\section{Experimental Methods}
\noindent\textbf{Experimental setup.} The simulation circuit is implemented on a $^{171}$Yb$^+$ ion confined in a micro-fabricated surface trap \cite{Revelle2020}. The qutrit energy levels are encoded as the hyperfine energy levels of the $^2$S$_{1/2}$ orbital: $\ket{0} \equiv \ket{F=0;m_F = 0}$, $\ket{1} \equiv \ket{F=1;m_F = 0}$, and $\ket{2} \equiv \ket{F=1;m_F = -1}$. In this ion, the $\ket{0}$ and $\ket{1}$ states remain coherent and form an ideal qubit state. The coherence of $\ket{2}$  depends strongly on the ambient magnetic field noise. We use a mu-metal shield to reduce the magnetic field noise experienced by the atomic ion by over two orders of magnitude, so all three qutrit states remain highly coherent~\cite{fang2023realization}. The transitions from $\ket{0}$ to $\ket{1}$ and $\ket{2}$ are achieved using stimulated Raman transitions driven with a pair of laser beams~\cite{wang2020}. We choose laser polarization settings so that the Rabi frequencies of $\ket{0}$ to $\ket{1}$ and $\ket{0}$ to $\ket{2}$ are close ($ 2\pi \times 17.30$ kHz and $2\pi \times17.49$ kHz, respectively). Acousto-optic modulators (AOMs) are used to tune the frequency and phase of each laser beam, while also allowing the beams to act as switches for each transition. At the end of the simulation, we have the ability to measure the probability of the qutrit in each of the states $\ket{0}, \ket{1}$, and $\ket{2}$. In a standard readout approach using a state-dependent fluorescence technique, the $\ket{0}$ state remains dark while both $\ket{1}$ and $\ket{2}$ states scatter photons (and therefore remain indistinguishable) upon illumination with the readout beam. To distinguish these two states, we first swap the population of the $\ket{1} (\ket{2})$ state with the $\ket{0}$ state using the Raman transition and then perform measurements of the dark state population to determine the probability of the qutrit state (prior to the swap) being in the $\ket{1}$ ($\ket{2}$) state. Details of the experimental setup are described in Ref. \cite{wang2020,fang2023realization}.  

\noindent\textbf{Trotterization and trapped-ion operation.} The Trotterization method provides a way to simulate the time evolution of a Hamiltonian with multiple non-commuting terms by simulating the unitary operation corresponding to each individual term for a short time duration, and repeating this for each term and time step~\cite{trotter1959product, suzuki1976generalized}. We use the following Trotterization method to simulate the time evolution of $\hat{H}_1(t)$, which can be applied straightforwardly  to $\hat{H}_2$. $\hat{H}_1(t)$ is written in the interaction picture as
\begin{equation}
    \hat{H}_{I,1}(t) = \hat{H}_{I,1}^{(1)}(t) + \hat{H}_{I,1}^{(2)}(t), \label{eqn:PLET1I}
\end{equation}
where
\begin{align}
    \hat{H}_{I,1}^{(1)}(t) &= \mu_1 E_x(t) e^{i(\omega_G - \omega_{D_1})t} \ket{G}\bra{D_1} + h.c., \label{eqn:PLET1I1}\\
    \hat{H}_{I,1}^{(2)}(t) &= \mu_2 E_y(t) e^{i(\omega_G - \omega_{D_2})t} \ket{G}\bra{D_2} + h.c. \label{eqn:PLET1I2}
\end{align}
The time evolution $\hat{U}$ with respect to $\hat{H}_{I,1}(t)$ up to time $T$ is Trotterized into $N$ discrete time steps. 
\begin{equation}
    \hat{U} = \hat{U}_N \hat{U}_{N-1} \cdots \hat{U}_1 = \hat{U}_{\rm ideal}(T) + O[(T/N)^3], \label{eqn:Trotterization}
\end{equation}
where 
\begin{equation}
    \hat{U}_{\rm ideal}(T) = \mathcal{T}\exp\left(-i\int_{0}^T H_{I,1}(t')dt'\right). 
\end{equation}
Here, $\mathcal{T}$ is the time-ordering operator, which orders the exponentiated Hamiltonians in chronological order. Also, the time evolution for each time step is built from a second-order Trotter formula~\cite{suzuki1976generalized} 
\begin{equation} \label{eqn:Trot}
    \hat{U}_j = e^{-i \hat{H}_{I,1}^{(1)}(t_j) \frac{T}{2N}}
    e^{-i \hat{H}_{I,1}^{(2)}(t_j) \frac{T}{N}}
    e^{-i \hat{H}_{I,1}^{(1)}(t_j) \frac{T}{2N}},
\end{equation}
where $t_j \equiv (j-1/2)T/N$. Each step corresponds to a simulated time evolution of $T/N$ and we plot the dynamics according to this simulated time. $N$ needs to be sufficiently large to simulate the time evolution of the system accurately. 

We map the molecular states $\ket{G}$, $\ket{D_1}$, and $\ket{D_2}$ to the trapped-ion qutrit states $\ket{0}$, $\ket{1}$, and $\ket{2}$, respectively. The evolution of the molecular states described above can be mapped to single-qutrit operations. For a transition between $\ket{0}$ and $\ket{\alpha}$ ($\alpha=1, 2$), the time evolution up to time $\tilde{\tau}$ is given by
\begin{equation}
   \tilde{U} = e^{-i \tilde{H} \tilde{\tau}},
\end{equation}
where
\begin{equation} \label{eqn:HTI}
    \tilde{H}_{\alpha} = \frac{\Omega_{\alpha}}{2} e^{i \phi_{\alpha}} \ket{0}\bra{\alpha} + h.c.,
\end{equation}
Here, $\Omega_{\alpha}$ and $\phi_{\alpha}$ are the Rabi frequency and phase, respectively, determined by the intensity and phase of the laser beam that drives the transition. Thus, the values of amplitude $\mu_1 E_x(t_j) T/2N$ [$\mu_2 E_y(t_j) T/N$] and phase $(\omega_G - \omega_{D_1})t_j$ [$(\omega_G - \omega_{D_2})t_j$] of each term in Eq.~(\ref{eqn:Trot}) can be mapped to $\Omega_\alpha \tilde{\tau}$ and $\phi_\alpha$ of the corresponding qutrit control operation, respectively. We can program these quantities by shaping the control laser beams.

In our experiments, it is desirable to use a fixed value of $\Omega_\alpha$ because of the difficulty associated with calibrating and stabilizing the laser intensity that drives the transition between the atomic states. Thus, instead of tuning $\Omega_\alpha$, we vary the evolution time $\tilde{\tau}$ for each Trotterization step with a constant Rabi frequency $\Omega_\alpha$ to simulate the electric field that varies over time. This method using Trotterization allows us to perform accurate simulations with a targeted upper bound on the error. 

\section*{Acknowledgments}
The authors express their gratitude to Ye Wang for his valuable contribution to the experimental setup. This research is funded by the Office of the Director of National Intelligence - Intelligence Advanced Research Projects Activity, through the ARO contract W911NF-16-1-0082, which provides support for the experimental apparatus utilized in Kim's laboratory, as well as by the DOE BES awards DE-SC0019400 and DE-SC0019449, which support Beratan's and Kim's laboratories, respectively, in terms of simulation methodology and analysis. In addition, the authors acknowledge the support of the NSF Quantum Leap Challenge Institute for Robust Quantum Simulation Grant No. OMA-2120757, which supports the Brown-Kim joint laboratory's experimental implementation.

\section{Supporting Information}

\subsection{Trapped-ion qubit simulations} \label{app:qubit}
In the main text we use a single trapped-ion qutrit to simulate the PLET dynamics. A more conventional experimental architecture than a qutrit system is a qubit system, where two atomic states per ion are used to encode the molecular electronic states. As $\hat{H}_1(t)$ and $\hat{H}_2$ each describe three states, at least two ions and two-qubit operations are required to map these Hamiltonians to trapped-ion qubits. Two-qubit operations require coupling the ions' internal qubit states to an ``entanglement bus'', which is the collective motion of the ion chain~\cite{CiracZoller, Molmer99, Sorensen99}. The normal modes of this collective motion, known as the motional modes, are more susceptible to external noise and thus often become the major limitation to realizing quantum computation and simulation with high accuracy~\cite{Cetina22, ff23}.

In this section, we briefly describe how the Hamiltonians $\hat{H}_1(t)$ and $\hat{H}_2$ in Eqs.~(\ref{eqn:PLET1}) and (\ref{eqn:PLET2}) can be simulated using single-qubit and two-qubit operations on two qubits. For brevity, we only show for the electron-transfer Hamiltonian $\hat{H}_2$, as it is straightforward to use the same method for the photo-excitation Hamiltonian $\hat{H}_1$. We first rewrite $\hat{H}_2$ in matrix form as 
\begin{equation} \label{eqn:PLET2mat}
    \hat{H}_{2} = 
    \begin{pmatrix}
    \tilde{\omega}_A & V_1 & V_2\\
    V_1 & \tilde{\omega}_{D_1} & 0 \\
    V_2 & 0 & \tilde{\omega}_{D_2}
    \end{pmatrix}, 
\end{equation}
where we define $\tilde{\omega}_j \equiv \omega_j - (\omega_{A} + \omega_{D_1} + \omega_{D_2})/3$ such that $\sum_j \tilde{\omega}_j = 0$ ($j \in \{A, D_1, D_2\}$), as redefining the zero-point energy does not change the dynamics with respect to the Hamiltonian. 

To simulate Eq.~(\ref{eqn:PLET2mat}) using two qubits, we map $\ket{A}, \ket{D_1}$ and $\ket{D_2}$ to the two-qubit computational basis states $\ket{00}, \ket{01}$ and $\ket{10}$, respectively. The $\ket{11}$ state needs to be decoupled from the evolution of the other three states. Thus, the corresponding two-qubit Hamiltonian is expressed in the matrix form as
\begin{equation} 
    \hat{H}_{2}^{\rm (qubit)} = 
    \begin{pmatrix}
    \tilde{\omega}_A & V_1 & V_2 & 0\\
    V_1 & \tilde{\omega}_{D_1} & 0 & 0 \\
    V_2 & 0 & \tilde{\omega}_{D_2} & 0 \\
    0 & 0 & 0 & 0 
    \end{pmatrix}.
\end{equation}
This trace-less Hamiltonian can be expressed as a linear sum of two-qubit Pauli matrices:
\begin{align} \label{eqn:PLET2qubit}
    \hat{H}_{2}^{\rm (qubit)} = &a_1 \hat{\sigma}_x\otimes \mathbf{I} + a_2 \mathbf{I} \otimes \hat{\sigma}_x + 
            a_3 \hat{\sigma}_z\otimes \mathbf{I} + a_4 \mathbf{I} \otimes \hat{\sigma}_z \nonumber \\
            &+ b_1 \hat{\sigma}_x\otimes \hat{\sigma}_z +
            b_2 \hat{\sigma}_z\otimes \hat{\sigma}_x + b_3 \hat{\sigma}_z\otimes \hat{\sigma}_z ,
\end{align}
where 
\begin{align*}
a_1 &= (\tilde{\omega}_A+\tilde{\omega}_{D_1})/2,\\
a_2 &= (\tilde{\omega}_A+\tilde{\omega}_{D_2})/2,\\
a_3 &= b_1 = V_2/2,\\
a_4 &= b_2 = V_1/2,\\
b_3 &= -(\tilde{\omega}_{D_1}+\tilde{\omega}_{D_2})/2.
\end{align*}

The time evolution of the system with the Hamiltonian of Eq.~(\ref{eqn:PLET2qubit}) can be simulated using Trotterization. Specifically, for each Trotterization time step, the evolution with respect to each of the first four terms can be simulated using single-qubit gates. Also, the evolution with respect to each of the last three terms of Eq.~(\ref{eqn:PLET2qubit}) can be simulated using the M{\o}lmer-S{\o}rensen~\cite{Molmer99, Sorensen99} interaction, which implements the Hamiltonian proportional to $\hat{\sigma}_x\otimes \hat{\sigma}_x$, conjugated with single-qubit Hadamard gates. 

\subsection{Qutrit vs. qubit implementation}\label{app:qutritvsqubit}

In this section, we use classical computers to compare the predicted accuracy of simulating the electron-transfer Hamiltonian $\hat{H}_2$ in Eq.~(\ref{eqn:PLET2}) using trapped-ion qubits and qutrit. For qubit simulations, single-qubit and two-qubit operations are performed on two ion qubits as described in Sec.~\ref{app:qubit}. For qutrit simulations, single-qutrit operations are performed on one ion qutrit. Since single-qutrit operations do not require interaction between different ions via the collective motion of the ions, the operations are not affected by the decoherence induced by the motional modes. Also, single-qutrit operations have a much shorter gate time than two-qubit operations, which reduces the impact of decoherence. As such, qutrit simulations are expected to achieve significantly higher accuracy than qubit simulations in predicting the time evolution of state populations. 

We use QuTip~\cite{Qutip} to numerically simulate the quantum simulation using qubits and qutrit under the relevant noise. Specifically, we solve the Lindblad master equation~\cite{lindblad1976generators}
\begin{equation*}
    \frac{d\hat{\rho}}{dt} = -i [\hat{H}, \hat{\rho}] + \sum_k \left(\hat{L}_k \hat{\rho} \hat{L}_k^\dagger - \frac{1}{2} \hat{L}_k^\dagger \hat{L}_k \hat{\rho} - \frac{1}{2} \hat{\rho} \hat{L}_k^\dagger \hat{L}_k\right),
\end{equation*}
where $\hat{\rho}$ is the density matrix, $\hat{H}$ is the Hamiltonian of the desired operation, and $\hat{L}_k$ is the $k$th Lindblad operator of the corresponding decoherence process.

\begin{figure}[ht]
\centering
\includegraphics[width=8.5cm]{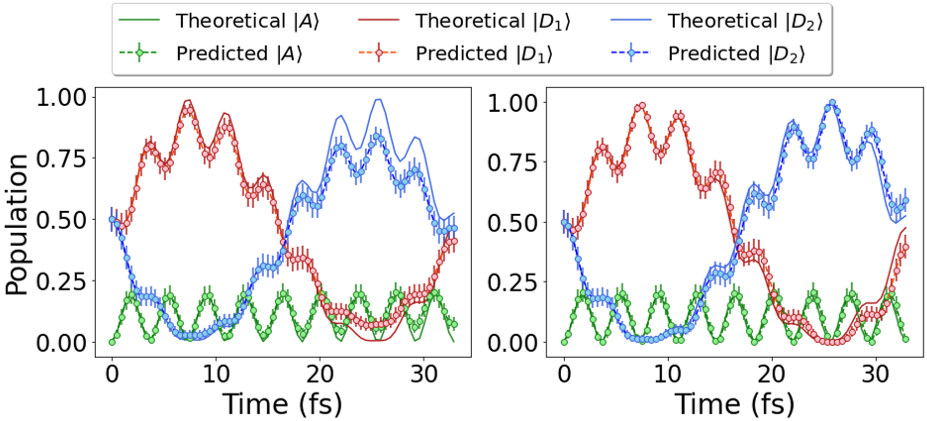}
\caption{Comparison between the qubit (left) and qutrit (right) simulations. The electron-transfer Hamiltonian $\hat{H}_2$ with parameter values of $\omega_{D_1} = \omega_{D_2} = 3.89$ eV, $\omega_{A} = 3.01$ eV, and $V_1 = V_2 = 0.25$ eV is simulated. The initial phase difference between $\ket{D_1}$ and $\ket{D_2}$ is set as $\phi = 90^{\circ}$. Solid lines represent the theoretically predicted state populations $P_{\rm th}$, and points represent the predictions of state populations obtained by quantum simulation $P_{\rm pred}$, predicted by the Lindblad master equation. 70 Trotter steps are used for the quantum simulation. The qubit case (left) shows larger deviations between $P_{\rm th}$ and $P_{\rm pred}$ compared to the qutrit case (right).}
\label{fig: dephasing}
\end{figure}

For predictions of the two-qubit quantum simulations, we simulate the evolution of the composite state of two qubits and two (out-of-phase and in-phase) motional modes in the direction perpendicular to the ion chain's axis. We assume that all single-qubit gates are instantaneous and perfectly accurate. Thus, all predicted experimental errors arise from the two-qubit operations, due to dephasing and heating of the motional modes. The motional dephasing is described by the Lindblad operator $\hat{L}_m = \sqrt{2/\tau_m} \hat{a}^\dagger \hat{a}$, where $\hat{a}$ ($\hat{a}^\dagger$) is the annihilation (creation) operator of the motional mode and $\tau_m$ is the motional coherence time. The motional heating is described by $\hat{L}_+ = \sqrt{\Gamma} \hat{a}^\dagger$ and $\hat{L}_- = \sqrt{\Gamma} \hat{a}$, where $\Gamma$ is the heating rate~\cite{wang2020, ff23}. We use $\tau_m = 8$ ms for both modes and $\Gamma = 10$ ($200$) $\rm{s}^{-1}$ for the out-of-phase (in-phase) motional mode, as characterized in our experimental setup. The two-qubit operations are assumed to be performed using the out-of-phase mode, which is less susceptible to heating. 

For predictions of the single-qutrit quantum simulations, we simulate the evolution of a 3-level atom. The dominant sources of errors are dephasing due to fluctuations of laser phase and magnetic field. Typically, magnetic field has negligible effects on the hyperfine qubit with $m_F=0$ states. However, when a qutrit is used, magnetic field fluctuations cause the dephasing of the Zeeman state $\ket{2} \equiv \ket{F=1;m_F = -1}$. Thus, $\ket{2}$ has a shorter coherence time than $\ket{1}$. The dephasing of $\ket{1}$ and $\ket{2}$ (with respect to $\ket{0}$) is described by the Lindblad operators $\hat{L}_1 = \sqrt{1/\tau_1} (\ket{0}\bra{0} - \ket{1}\bra{1})$ and $\hat{L}_2 = \sqrt{1/\tau_2} (\ket{0}\bra{0} - \ket{2}\bra{2})$, respectively. We use $\tau_1 = 300$ ms and $\tau_2 = 75$ ms, as characterized in our experimental setup. 

We denote the state populations predicted by this numerical simulation using the Lindblad master equation as $P_{\rm pred}$. Figure~\ref{fig: dephasing} illustrates a comparison between the values of $P_{\rm pred}$ and the theoretical prediction values of $P_{\rm th}$. The number of Trotterization steps is set as 70 for both qubit and qutrit simulations. The simulation time for the time evolution of $\hat{H}_2$ with qutrits is significantly shorter, at $0.226$ ms, compared to the qubit case which takes $17.92$ ms. This is due to the higher Rabi frequency of single-qutrit gates ($2\pi \times 17.30, \: 17.49$ kHz) compared to the motional-sideband Rabi frequency ($2\pi \times 3.92, \: 5.03$ kHz) that is relevant to two-qubit gates. Also, the qutrit states' coherence time $\tau_1$ and $\tau_2$ are significantly longer than the motional coherence time $\tau_m$ that limits the accuracy of qubit simulations. Thus, as shown in Fig.~\ref{fig: dephasing}, $P_{\rm pred}$ match $P_{\rm th}$ significantly more closely in qutrit simulations than qubit simulations, due to the longer relevant coherence time and shorter simulation time. 

When simulating larger systems with more than three states, entangling gates will also be needed in the qutrit case. However, the number of required entangling gates can be fewer than that in the case of qubits~\cite{brylinski2002universal,gokhale2019asymptotic}. Hence, employing a smart encoding platform, such as qutrits, may achieve simulations of higher accuracy. 

\bibliographystyle{unsrt}  
\bibliography{references} 

\begin{thebibliography}{10}

\bibitem{etmarcus}
R.~A. Marcus.
\newblock Chemical and electrochemical electron-transfer theory.
\newblock {\em Annu. Rev. Phys. Chem}, 15, 1964.

\bibitem{etdb}
D.~N. Beratan.
\newblock Why are dna and protein electron transfer so different?
\newblock {\em Annu. Rev. Phys. Chem}, 70, 2019.

\bibitem{prytkova2007coupling}
T.~R Prytkova, I.~V Kurnikov, and D.~N. Beratan.
\newblock Coupling coherence distinguishes structure sensitivity in protein
  electron transfer.
\newblock {\em Science}, 315(5812):622--625, 2007.

\bibitem{spiroscpl}
S.~S. Skourtis, D.~N. Beratan, R.~Naaman, A.~Nitzan, and D.~H. Waldeck.
\newblock Chiral control of electron transmission through molecules.
\newblock {\em Phys. Rev. Lett.}, 101, 2008.

\bibitem{ray1999asymmetric}
K~Ray, S.~P. Ananthavel, D.~H. Waldeck, and R.~Naaman.
\newblock Asymmetric scattering of polarized electrons by organized organic
  films of chiral molecules.
\newblock {\em Science}, 283(5403):814--816, 1999.

\bibitem{yang2013circularly}
Y.~Yang, R.~C. Da~Costa, M.~J. Fuchter, and A.~J. Campbell.
\newblock Circularly polarized light detection by a chiral organic
  semiconductor transistor.
\newblock {\em Nat. Photon.}, 7(8):634--638, 2013.

\bibitem{feynman2018simulating}
R.~P. Feynman.
\newblock Simulating physics with computers.
\newblock In {\em Feynman and computation}, pages 133--153. CRC Press, 2018.

\bibitem{Maier19}
C.~Maier, T.~Brydges, P.~Jurcevic, N.~Trautmann, C.~Hempel, B.~P. Lanyon,
  P.~Hauke, R.~Blatt, and C.~F. Roos.
\newblock Environment-assisted quantum transport in a 10-qubit network.
\newblock {\em Phys. Rev. Lett.}, 122:050501, Feb 2019.

\bibitem{Gorman18}
D.~J. Gorman, B.~Hemmerling, E.~Megidish, S.~A. Moeller, P.~Schindler,
  M.~Sarovar, and H.~Haeffner.
\newblock Engineering vibrationally assisted energy transfer in a trapped-ion
  quantum simulator.
\newblock {\em Phys. Rev. X}, 8:011038, Mar 2018.

\bibitem{trotter1959product}
H.~F. Trotter.
\newblock On the product of semi-groups of operators.
\newblock {\em Proceedings of the American Mathematical Society},
  10(4):545--551, 1959.

\bibitem{suzuki1976generalized}
M.~Suzuki.
\newblock Generalized trotter's formula and systematic approximants of
  exponential operators and inner derivations with applications to many-body
  problems.
\newblock {\em Commun. Math. Phys.}, 51(2):183--190, 1976.

\bibitem{hughes1996decoherence}
R.~J. Hughes, D.~F.~V. James, E.~H. Knill, R.~Laflamme, and A.~G. Petschek.
\newblock Decoherence bounds on quantum computation with trapped ions.
\newblock {\em Phys. Rev. Lett.}, 77(15):3240, 1996.

\bibitem{Klimov03}
A.~B. Klimov, R.~Guzm\'an, J.~C. Retamal, and C.~Saavedra.
\newblock Qutrit quantum computer with trapped ions.
\newblock {\em Phys. Rev. A}, 67:062313, Jun 2003.

\bibitem{low2020practical}
P.~Low, B.~M. White, A.~A. Cox, M.~L. Day, and C.~Senko.
\newblock Practical trapped-ion protocols for universal qudit-based quantum
  computing.
\newblock {\em Phys. Rev. Res.}, 2(3):033128, 2020.

\bibitem{ringbauer2022universal}
M.~Ringbauer, M.~Meth, L.~Postler, R.~Stricker, R.~Blatt, P.~Schindler, and
  T.~Monz.
\newblock A universal qudit quantum processor with trapped ions.
\newblock {\em Nat. Phys.}, 18(9):1053--1057, 2022.

\bibitem{hrmo2022native}
P.~Hrmo, B.~Wilhelm, L.~Gerster, M.~W. van Mourik, M.~Huber, R.~Blatt,
  P.~Schindler, T.~Monz, and M.~Ringbauer.
\newblock Native qudit entanglement in a trapped ion quantum processor.
\newblock {\em arXiv:2206.04104}, 2022.

\bibitem{chi2022programmable}
Y.~Chi, J.~Huang, Z.~Zhang, J.~Mao, Z.~Zhou, X.~Chen, C.~Zhai, J.~Bao, T.~Dai,
  H.~Yuan, et~al.
\newblock A programmable qudit-based quantum processor.
\newblock {\em Nat. Commun.}, 13(1):1166, 2022.

\bibitem{cervera2022experimental}
A.~Cervera-Lierta, M.~Krenn, A.~Aspuru-Guzik, and A.~Galda.
\newblock Experimental high-dimensional greenberger-horne-zeilinger
  entanglement with superconducting transmon qutrits.
\newblock {\em Phys. Rev. Appl.}, 17(2):024062, 2022.

\bibitem{Potocnik18}
A.~Poto{\v{c}}nik, A.~Bargerbos, F.~A. Y.~N. Schr{\"o}der, S.~A. Khan, M.~C.
  Collodo, S.~Gasparinetti, Y.~Salath{\'e}, C.~Creatore, C.~Eichler, H.~E.
  T{\"u}reci, A.~W. Chin, and A.~Wallraff.
\newblock Studying light-harvesting models with superconducting circuits.
\newblock {\em Nat. Commun.}, 9(1):904, Mar 2018.

\bibitem{Wang18}
B.~Wang, M.~Tao, Q.~Ai, T.~Xin, N.~Lambert, D.~Ruan, Y.~Cheng, F.~Nori,
  F.~Deng, and G.~Long.
\newblock Efficient quantum simulation of photosynthetic light harvesting.
\newblock {\em Npj Quantum Inf.}, 4(1):52, Oct 2018.

\bibitem{MacDonell21}
R.~J. MacDonell, C.~E. Dickerson, C.~J.~T. Birch, A.~Kumar, C.~L. Edmunds,
  M.~J. Biercuk, C.~Hempel, and I.~Kassal.
\newblock Analog quantum simulation of chemical dynamics.
\newblock {\em Chem. Sci.}, 12:9794--9805, 2021.

\bibitem{whitlow2022simulating}
J.~Whitlow, Z.~Jia, Y.~Wang, C.~Fang, J.~Kim, and K.~R. Brown.
\newblock Simulating conical intersections with trapped ions.
\newblock {\em arXiv:2211.07319}, 2022.

\bibitem{valahu2022direct}
C.~H. Valahu, V.~C. Olaya-Agudelo, R.~J. MacDonell, T.~Navickas, A.~D. Rao,
  M.~J. Millican, J.~B. P{\'e}rez-S{\'a}nchez, J.~Yuen-Zhou, M.~J Biercuk,
  C.~Hempel, et~al.
\newblock Direct observation of geometric phase in dynamics around a conical
  intersection.
\newblock {\em arXiv:2211.07320}, 2022.

\bibitem{wang2023observation}
C.~S. Wang, N.~E. Frattini, B.~J. Chapman, S.~Puri, S.~M. Girvin, M.~H.
  Devoret, and R.~J. Schoelkopf.
\newblock Observation of wave-packet branching through an engineered conical
  intersection.
\newblock {\em Phys. Rev. X.}, 13(1):011008, 2023.

\bibitem{Revelle2020}
M.~C. Revelle.
\newblock Phoenix and peregrine ion traps.
\newblock {\em arXiv:2009.02398}, 2020.

\bibitem{fang2023realization}
C.~Fang, Y.~Wang, K.~Sun, and J.~Kim.
\newblock Realization of scalable cirac-zoller multi-qubit gates.
\newblock {\em arXiv:2301.07564}, 2023.

\bibitem{wang2020}
Y.~Wang, S.~Crain, C.~Fang, B.~Zhang, S.~Huang, Q.~Liang, P.~H. Leung, K.~R.
  Brown, and J.~Kim.
\newblock High-fidelity two-qubit gates using a
  microelectromechanical-system-based beam steering system for individual qubit
  addressing.
\newblock {\em Phys. Rev. Lett.}, 125:150505, Oct 2020.

\bibitem{CiracZoller}
J.~I. Cirac and P.~Zoller.
\newblock Quantum computations with cold trapped ions.
\newblock {\em Phys. Rev. Lett.}, 74:4091--4094, May 1995.

\bibitem{Molmer99}
K.~M{\o}lmer and A.~S{\o}rensen.
\newblock Multiparticle entanglement of hot trapped ions.
\newblock {\em Phys. Rev. Lett.}, 82(9):1835, 1999.

\bibitem{Sorensen99}
A.~S{\o}rensen and K.~M{\o}lmer.
\newblock Quantum computation with ions in thermal motion.
\newblock {\em Phys. Rev. Lett.}, 82(9):1971, 1999.

\bibitem{Cetina22}
M.~Cetina, L.~N. Egan, C.~Noel, M.~L. Goldman, D.~Biswas, A.~R. Risinger,
  D.~Zhu, and C.~Monroe.
\newblock Control of transverse motion for quantum gates on individually
  addressed atomic qubits.
\newblock {\em PRX Quantum}, 3:010334, Mar 2022.

\bibitem{ff23}
M.~Kang, Y.~Wang, C.~Fang, B.~Zhang, O.~Khosravani, J.~Kim, and K.~R. Brown.
\newblock Designing filter functions of frequency-modulated pulses for
  high-fidelity two-qubit gates in ion chains.
\newblock {\em Phys. Rev. Appl.}, 19:014014, Jan 2023.

\bibitem{Qutip}
J.R. Johansson, P.D. Nation, and Franco Nori.
\newblock Qutip: An open-source python framework for the dynamics of open
  quantum systems.
\newblock {\em Comput. Phys. Commun.}, 183(8):1760--1772, 2012.

\bibitem{lindblad1976generators}
G.~Lindblad.
\newblock On the generators of quantum dynamical semigroups.
\newblock {\em Commun. Math. Phys.}, 48:119--130, 1976.

\bibitem{brylinski2002universal}
J.~Brylinski and R.~Brylinski.
\newblock Universal quantum gates.
\newblock In {\em Mathematics of quantum computation}, pages 117--134. Chapman
  and Hall/CRC, 2002.

\bibitem{gokhale2019asymptotic}
P.~Gokhale, J.~M. Baker, C.~Duckering, N.~C. Brown, K.~R. Brown, and F.~T.
  Chong.
\newblock Asymptotic improvements to quantum circuits via qutrits.
\newblock In {\em Proceedings of the 46th International Symposium on Computer
  Architecture}, pages 554--566, 2019.

\end{thebibliography}

\end{document}